\documentclass[aps,showpacs,amsmath,nofootinbib,12pt]{revtex4}

\usepackage{amsfonts}
\usepackage{amssymb}
\usepackage{color}
\usepackage{mathrsfs}

\def\bc{\begin{center}}

\def\ec{\end{center}}
\def\be{\begin{eqnarray}}
\def\ee{\end{eqnarray}}

\newcommand{\omits}[1]{}

\definecolor{dyellow}{rgb}{1.,0.8,.0}
\definecolor{myblue}{rgb}{.1,.1,.7}
\definecolor{dcyan}{rgb}{.0,.6,.6}
\definecolor{dmagenta}{rgb}{0.6,0.0,0.6}
\definecolor{brown}{rgb}{0.6,0.2,0.}
\definecolor{darkblue}{rgb}{.0,.0,0.5}
\definecolor{darkred}{rgb}{0.75,0.0,0.0}
\definecolor{orange}{rgb}{1.,.6,.0}
\definecolor{dorange}{rgb}{0.8,.4,.0}
\definecolor{lightgray}{rgb}{0.7,0.7,0.7}
\definecolor{darkgreen}{rgb}{0.0,0.6,0.0}
\definecolor{purple}{rgb}{.4,.0,.4}

\def\blue{\color{blue}}

\def\delete{\color{lightgray}}

\def\eps{\epsilon}

\def\d#1#2{\frac{\displaystyle #1}{\displaystyle #2}}

\newcommand{\dS}{$d{S}$}
\newcommand{\Mink}{${M}ink$}

\newcommand{\dSR} {$d{S}$ special relativity}

\newcommand\btd{\raise 2pt
\hbox{$\hat\bigtriangledown$}\hskip 1.5pt}
\newcommand\bt{\raise 2pt
\hbox{$\bigtriangledown$}\hskip 1.5pt}

\def\cA{{\cal A}}
\def\cB{{\cal B}}

\def\PRD{{\it Phys. Rev.}~{\bf D}}

\def\PLA{{\it Phys. Lett.}~{\bf A}}

\def\no{\noindent}

\begin{document}

\title{From the Complete Yang Model to Snyder's Model,  de Sitter Special
Relativity and Their Duality\footnote{ This work is partly supported
by NSFC (Grant Nos 10701081 and 10775140), NKBRPC(2004CB318000),
Beijing Jiao-Wei Key project(KZ200810028013) {and Knowledge
Innovation Funds of CAS (KJCX3-SYW-S03)} }.}

\author{{Hong-Tu Wu}$^{1}$}
\email{lobby_wu@yahoo.com.cn}
\author{{Chao-Guang Huang}$^{2}$}
\email{huangcg@mail.ihep.ac.cn}
\author{{Han-Ying Guo}$^{3}$}
\email{hyguo@itp.ac.cn}

\affiliation{%
${}^1$Department of Mathematics, Capital Normal University, Beijing
100037, China.}
\affiliation{%
${}^2$  Institute of High Energy Physics, Chinese Academy of
Sciences, Beijing 100049, China,}
\affiliation{%
${}^3$ Institute of Theoretical Physics, Chinese Academy of
Sciences, Beijing 100080, China,}

\begin{abstract}
By means of Dirac procedure, we re-examine Yang's quantized
space-time model, its relation to Snyder's model, the \dS\ special
relativity and their  UV-IR duality. Starting from a dimensionless
\dS$_5$-space  in a 5+1-d \Mink-space a complete Yang model at both
classical and quantum level can be presented and there really exist
Snyder's model, the \dS\ special relativity and the duality.
\medskip

\no Keywords: Yang model, Snyder's model, de Sitter
special relativity, duality
\end{abstract}
\maketitle




Recently, based on Yang's quantized space-time model \cite{Yang}, we
present a complete Yang model \cite{YMDual} from a 5+1-d space with
\Mink-signature mainly by projective geometry method. We have shown
that in the  model, there are Snyder's model \cite{Snyder}, the \dS\
special relativity \cite{Lu,LZG,BdS,BdS05,IWR,T,NH,Lu05,C3, Duality,
Yan, PoI, Lu80} and their UV-IR duality \cite{Duality}.

 In this letter, we re-examine the same topic from a slightly different angle:
 Dirac procedure for the constrained systems \cite{Dirac}.
 Starting from a dimensionless \dS$_5$-space of
radius $R/a$ embedded in  a dimensionless 5+1-d \Mink-space, by
means of Dirac procedure, Yang's algebra \omits{\delete under Dirac
bracket }can be given at both classical and quantum level. We
\omits{\delete also} show that Snyder's model, the \dSR\ and  their
duality  can also be given from such a complete Yang model in a
slightly different way with Dirac procedure.


The dimensionless \dS$_5$-space reads 
\be\label{eq:quard}%
\mathscr{H}: && \eta_{\cA \cB}^{} \zeta^{\cA} \zeta^{\cB} =
-\frac{R^2}{a^2}, \ \cA,\cB=0,1, \cdots, 5,\\
&& d\chi^2=\eta_{\cA \cB}^{} d\zeta^{\cA} d\zeta^{\cB}
\ee%
with $\eta_{\cA \cB}^{}={\rm diag}(+, -, -, -, -, -)$,
where $a$ and $R$ are of length dimension.\omits{
$\zeta^\mu\,(\mu=0,1,2,3)$ in this equation corresponds to
$\eta_\mu$, $\zeta^4$ to $\eta$ and $\zeta^5$ to $\zeta$,
respectively.}

For a fictitious particle moving on $\mathscr{H}$ with a Lagrangian
function 
\begin{equation}
{\cal L} = -\sqrt{\eta_{\cA \cB}^{} \dot{\zeta}^\cA
\dot{\zeta}^\cB}, \quad \dot{\zeta}^\cA=\frac{d \zeta^\cA}{d\chi},
\label{lagrangian}
\end{equation}
 the `canonical momentum' and the 6-d `angular momentum' read,
respectively%
 \be\label{cmomentum} N_{\cA}=\frac{\partial
\mathcal{L}}{\partial
\dot{\zeta}^\cA}=-\eta_{\cA\cB}\frac{d \zeta^{\cB}}{d\chi},\\
 \mathscr{L}^{\cal AB}=\zeta^\cB N^\cA - \zeta^\cA
N^\cB. \label{angular momentum} \ee%
 Thus, there is a `phase space' with ($\zeta^{\cA}$, $N_{\cA}$)
\omits{span a `phase space' with} and Poisson bracket
\begin{equation}
\{f, g\}=\frac{\partial f}{\partial \zeta^{\cB}}\frac{\partial
g}{\partial N_{\cB}}-\frac{\partial f}{\partial N_{\cB
}}\frac{\partial g}{\partial \zeta^{\cB}}. \label{Poisson}
\end{equation}
The Hamiltonian  $H=N_\cB \dot{\xi}^\cB-\mathcal{L}$ {for the
particle vanishes.}

Note that there are two primary constraints: $\phi_1=\eta_{\cal
AB}\zeta^{\cA} \zeta^{\cB}+ \d {R^2}{a^2} =0$ from  Eq.
(\ref{eq:quard}), and $\phi_2=\eta^{\cal AB} N_\cA N_\cB-1=0$
as the `mass-shell' condition from Eqs. (\ref{lagrangian}) and
(\ref{cmomentum}). According to Dirac \cite{Dirac},  the total
Hamiltonian can be introduced as $H_T=H+\mu^m \phi_m, (m=1,2),$ and
the dynamical consistency condition leads to
\begin{equation}
\begin{split}
0&=\dot{\phi}_1=\{ \phi_1, H_T \}=4 \mu^2 N_\cB \zeta^\cB, \\
0&=\dot{\phi}_2=\{ \phi_2, H_T \}=-4 \mu^1 N_\cB \zeta^\cB.\
\end{split}
\end{equation}
Let $\phi_3=N_\cB \zeta^\cB=0$, then we get a secondary constraint
$\phi_3$. Its dynamical consistency condition {results in} $\mu^2= -
\d {R^2} {a^2} \mu^1$. Thus, the total Hamiltonian becomes
$H_T=\mu^1\left( \phi_1 -\d {R^2} {a^2} \phi_2 \right)$. Set
$\phi_0=\phi_1 - \d {R^2} {a^2} \phi_2=0$. Obviously, $\phi_0$ is of
the first class and  satisfies  the consistency condition. Hence,
there are one first-class constraint $\phi_0$ and two
second-class constraints $\phi_1,\; \phi_3$ with Dirac
bracket£º
\begin{equation}
\{f, g\}_{DB} :=\{f,g\}-\{ f, \phi_m\} C_{mk}^{-1}\{\phi_k,g\},
\ m,k=1,3,
\end{equation}
where $C$ is a reversible  matrix{, defined by}
\begin{equation}
C= 2 \d {R^2} {a^2}\left(
\begin{array}{cc}
0 & -1\\
1 & 0
\end{array}
\right).
\end{equation}
The basic non-vanishing  Dirac brackets with Jacobi identity read
\begin{equation}
\begin{split}
\{ N^\cA, N^\cB\}_{DB}&=-\d {a^2}{R^2}\mathscr{L}^{\cal AB}, \\
\{\zeta^\cA, N_\cB\}_{DB}&=\delta^\cA_\cB+\d {a^2}{R^2} \zeta^{\cA}
\zeta_{\cB}.
\end{split}
\end{equation}
Further, the  6-d `angular-momentum' $\mathscr{L}^{\cal AB}$ forms
an
$\mathfrak{so}(1,5)$ algebra 
\begin{equation}
\begin{split}
& \{{\mathscr L}^{\cal AB},
{\mathscr L}^{\cal CD}\}_{DB}\\
=&\eta^{\cal AD}{\mathscr L}^{\cal
BC}+\eta^{\cal BC}{\mathscr L}^{\cal AD}-\eta^{\cal AC}{\mathscr
L}^{\cal BD}-\eta^{ \cal BD} {\mathscr L}^{\cal AC},\label{so15}
\end{split}
\end{equation}
and is conserved\omits{%
 \begin{equation} \d d {d \chi}
\mathscr{L}^{\cal AB}=0,
\end{equation}}
under the `canonical equations'
\begin{eqnarray}
\dot{\zeta}^\cA &=& \{ \zeta^\cA, H_T \}_{DB}= - 2 \mu^1 \d
{R^2}{a^2} \eta^{\cal AB} N_\cB, \label{eq zeta}\\
\dot{N}_\cA &=& \{ N_\cA, H_T \}_{DB}= -2 \mu^1 \eta_{\cal
AB}\zeta^\cB. \label{eq N}
\end{eqnarray}
From the definition of $N_\cA$, it follows $\mu^1=\d {a^2}{2 R^2}$.

Introduce a set of variables ($c=1$)\footnote{$\epsilon_{ijk}$ is an
anti-symmetric tensor with $\epsilon_{123}=1$. The index  is raised
by $\eta^{ij}$, $i,j=1,2,3$.}:
\begin{equation}
\begin{split}
&{x}_0 = - a(\zeta^5 {N}_0+\zeta^0 {N}_5)=a\mathscr{L}^{50} , \\
&{x}_i  = a(\zeta^i {N}_5-\zeta^5 {N}_i) =-a \mathscr{L}^{5i}, \\
&{p}_0 =-\frac{\hbar}{R}(\zeta^4 {N}_0+\zeta^0 {N}_4)=\frac \hbar R
\mathscr{L}^{40},\\
&{p}_i  =\frac{\hbar}{R}(\zeta^i {N}_4-\zeta^4 {N}_i)=-\frac \hbar R \mathscr{L}^{4i}, \\
&{M}_i =-\hbar(\zeta^0 {N}_i+\zeta^i N_0)=-\hbar\mathscr{L}^{0i}, \\
&{L}_i  = - \hbar \eps_{ijk}\left(\zeta^j {N}^k\right)=\frac 1 2
\hbar \eps_{ijk}\mathscr{L}^{jk}, \\
&\psi  = \frac a R(\zeta^4 N^5-\zeta^5N^4)=\d a R \mathscr{L}^{45}.
\end{split} \label{yang variable}
\end{equation}
They satisfy Yang's algebra under Dirac bracket\footnote{This
$\psi$ is slightly different from Yang's $\xi$ \cite{Yang} by a
factor of length dimension. It doesn't affect the issue in
substance.}:
\begin{equation}
\begin{split}
& \{p^\mu, p^\nu\}_{DB}=\frac{\hbar}{R^2}l^{\mu\nu},  \\
& \{l^{\mu\nu}, p^\rho\}_{DB}=\eta^{\nu \rho}p^\mu-\eta^{\mu\rho}
p^\nu,\\
& \{x^\mu, x^\nu\}_{DB}=a^2 \hbar^{-1} l^{\mu\nu}, \\
& \{l^{\mu\nu}, x^\rho\}_{DB} =\eta^{\nu \rho} x^\mu -\eta^{\mu \rho}
x^\nu,\\
& \{x^\mu, p^\nu\}_{DB}=\hbar\eta^{\mu\nu}\psi,  \\
& \{\psi, x^\mu \}_{DB}=-a^2 \hbar^{-1} p^\mu,\\
& \{\psi, p^\mu \}_{DB}=\frac{\hbar^2}{R^2} x^\mu , \\
& \{\psi, l^{\mu \nu} \}_{DB} = 0, \\
& \{l^{\mu\nu},
l^{\rho\sigma}\}_{DB}=\eta^{\mu\sigma}l^{\nu\rho}+\eta^{
\nu\rho}l^{\mu\sigma}-\eta^{\mu\rho}l^{\nu\sigma}-\eta^{
\nu\sigma}l^{\mu\rho},
\end{split} \label{yang poisson}%
\end{equation}
with $x^\mu=\eta^{\mu\nu}x_\nu$, $p^\mu=\eta^{\mu\nu}p_\nu, ~\mu,
\nu = 0, \cdots 3$, $l^{0i}=-M_i$, and $\d 1 2 \eps_{ijk}l^{jk} =
L_i$.

Under the `canonical quantization', i.e. replacing $\zeta^{\cA}$ and
$N_{\cB}$ with the corresponding operators and the Dirac bracket
$\{\ ,\ \}_{DB}$ with $- i [\ ,\ ]$,  the quantized  commutator of
$\hat{\zeta}^{\cA} \; \mbox{and} \; \hat{N}_{\cB}$ reads
\begin{equation}
[\hat{\zeta}^{\cA}, \hat{N}_{\cB}]=i( \delta^{\cA}_{\cB}+ \d {a^2}
{R^2} \hat{\zeta}^{\cA} \hat{\zeta}_{\cB}).
\end{equation}
In the `coordinate picture',
\begin{equation}
\hat{N}_{\cA}=-i \left( \frac{\partial }{\partial \zeta^{\cA}}+\d
{a^2}{R^2} \zeta_{\cA} \zeta^{\cB} \d {\partial }{\partial
\zeta^{\cB}} \right),
\end{equation}
and an $\mathfrak{so}(1,5)$ in commutator for the 6-d
`angular-momentum' follows
\begin{equation}
\begin{split}
&[\hat{\mathscr{L}}^{\cal AB}, \hat{\mathscr{L}}^{\cal
CD}]\\
=&\eta^{\cal AD}\hat{{\mathscr L}}^{\cal BC}+\eta^{\cal
BC}\hat{{\mathscr L}}^{\cal AD}-\eta^{\mathscr AC}\hat{{\mathscr
L}}^{\cal BD}-\eta^{ \cal BD}\hat{ {\mathscr L}}^{\cal AC}.
\end{split}
\end{equation}
Obviously, Yang's algebra can be derived by replacing the variables
in Eqs. (\ref{yang variable}) with  corresponding operators.

 Note that in
Yang's algebra at both the classical and quantum level there are two
\dS\ subalgebras for $x^\mu$ (or $\hat{x}^\mu$) and  $p^\mu$ (or
$\hat{p}^\mu$) with a common $\mathfrak{so}(1,3)$ for $l^{\mu\nu}$
(or $\hat{l}^{\mu\nu}$), respectively. There is an important
property: Yang's algebra with respect to the 6-d `angular momentum'
(\ref{angular momentum}) is invariant under a ${\bf Z}_2=\{e,\tau\
|\tau^2=e\}$ \cite{YMDual} with $\tau$
\begin{equation}
\tau:\quad a \ \leftrightarrow \ \d \hbar R, \quad x^\mu \
\leftrightarrow \ p^\mu, \quad \psi \ \leftrightarrow \ -\psi .
\label{duality}
\end{equation}
If the universal constants $a$ and $R$ are of the Planck length
$\ell_p$ and the cosmological radius $R=\sqrt{3/\Lambda}$, the ${\bf
Z}_2$-duality is a UV-IR duality.


As was shown in \cite{YMDual}, it is obvious that \dS$_5$-space
$\mathscr{H}$ contains two \dS\ sub-manifolds and both Snyder's
model and the \dSR\ can be derived from the complete Yang model in
$\mathscr{H}$. In the subspace ${\mathscr I}_1$ of $\mathscr{H}\subset
\mathscr{M}^{1,5}$ as an intersection %
\be\label{Ia}%
{\mathscr I}_1=\mathscr{H}|_{\zeta^4=0}:\quad \mathscr{H} \cap
\mathscr{P}_1\subset \mathscr{M}^{1,5},\ee%
where $\mathscr{P}_1$ is the hyperplane defined by $\zeta^4=0$,
let
\begin{equation}
\eta_\mu=\frac{\hbar}{R} \zeta^\mu \quad \mbox{and} \quad
\eta_4=\frac{\hbar}{R} \zeta^5, \label{YStrans}
\end{equation}
then a \dS-space of momentum  with $\eta^{AB}={\rm diag}(+,-,-,-,-)$,
$A, B =0,\ldots, 4$, follows
\begin{equation}\label{eq:momentum}
\begin{split}
{\cal H}_a:~~& \eta^{AB} \eta_A^{}\eta_B^{}=-\frac{\hbar^2}{a^2},\\
&ds_a^2=\eta^{AB}\eta_A^{}\eta_B^{}=\frac{\hbar^2}{R^2} d^2 \chi.
\end{split}\end{equation}
In the subspace $\mathscr{I}_1$ with the variables (\ref{YStrans}),
the operators $\hat{x}^0, \; \hat{x}^i, \;
\hat{M}_i, \; \mbox{and} \; \hat{L}_i$ in Yang's algebra become
\begin{equation}
\begin{split}
\hat{x}_0  = ia \left ( \eta_4^{}\d{\partial }{\partial \eta_0^{}} +
\eta_0^{}\d{\partial }{\partial \eta_4} \right), &\  \hat{x}_i =
ia \left ( \eta_4^{}\d{\partial }{\partial \eta_i}
-\eta_i^{}\d{\partial }{\partial \eta_4} \right),\\
\hat{M}_i  = i\hbar \left ( \eta_0^{}\d{\partial }{\partial \eta_i^{}} +
\eta_i^{}\d{\partial }{\partial \eta_0} \right), & \  \hat{L}_i
\omits{ = \d 1 2 i\hbar \eps_{ijk} \left ( \eta^j \d{\partial
}{\partial \eta_k}- \eta^k \d{\partial }{\partial \eta_j} \right)}=
i \hbar \eps_{ijk} \left ( \eta^j \d {\partial }{\partial \eta^k}
\right),
\end{split}
\end{equation}
which form an $\mathfrak{so}(1,4)$  under Lie bracket and keep
Snyder's \dS-space of momentum (\ref{eq:momentum}) invariant. Thus,
Snyder's model is contained in the complete Yang model.

On the other hand, in the subspace ${\mathscr I}_2$ of $\mathscr{H}\subset
\mathscr{M}^{1,5}$ as another intersection %
\omits{in the intersection $\mathscr{I}_2$ or the
hyper-surface $\mathscr{P}_2$ of the space $\mathscr{H}$}
\begin{equation}
\mathscr{I}_2= \mathscr{H}|_{\zeta^5=0}: \quad  \quad
\mathscr{H} \cap \mathscr{P}_2 \subset \mathscr{M}^{1,5},
\end{equation}
where $\mathscr{P}_2$ is the hyperplane defined by $\zeta^5=0$, let
\begin{equation}
\xi^\mu=a \zeta^\mu \quad \mbox{and} \quad \xi^4=a \zeta^4,
\label{dstran}
\end{equation}
then a \dS-spacetime follows
\begin{equation}\label{eq:coordinate}
\begin{split}
{\cal H}_R:&\quad \eta_{AB} \xi^A \xi^B = -R^2,\\
& ds_R^2=\eta_{AB}
d\xi^A d\xi^B=a^2 d^2 \chi.
\end{split}\end{equation}
In the subspace $\mathscr{I}_2$ with the variables
(\ref{dstran}), the ${\cal A,\;B} \neq 5$ components of the 6-d
`angular momentum' consist of  the 5-d angular momentum in the \dSR.
Now, the classical variables $p_0,\; p_i, \; M_i, \; \mbox{and} \;
L_i$ defined in (\ref{yang variable}) become
\begin{equation}
\begin{split}
&p_0 = \d 1 R \d \hbar a \left( \xi^4 \d {d \xi^0 }{ds_R^{}} - \xi^0
\d {d \xi^4 }{ds_R^{}} \right), \\
& \ p_i = \d 1 R \d \hbar a
\left( \xi^i \d {d \xi^4 }{ds_R^{}}-\xi^4
\d {d \xi^i }{d s_R} \right),\\
&M_i = \d \hbar a \left( \xi^i \d {d \xi^0 }{ds_R^{}}-\xi^0 \d {d
\xi^i }{d s_R} \right),\\
& \ L_i = \d \hbar a \eps_{ijk} \left(
\xi^j \d {d \xi^k}{d s_R} \right),
\end{split}
\end{equation}
which are the 5-d formalism of the conserved quantities for a free
particle with  mass $\d \hbar a$ in the \dSR. Thus, the \dSR\ is
also contained in the complete Yang model.

As for the duality between Snyder's model and the \dSR\
\cite{Duality}, it can also be studied in the complete Yang model.
In fact, the \dS$_4$-spaces for both Snyder's model and the \dSR\
are given by the sub-manifolds with $\zeta^4=0$ and $\zeta^5=0$,
respectively, from the \dS$_5$-space $\mathscr{H}$ in the  model.
Therefore, the duality is related to a one-to-one correspondence
between $\zeta^4$ and $\zeta^5$. Since these  variables are only
related to $\psi$ (or $\zeta$ in \cite{Yang}), and $\psi$ becomes
$-\psi$ when $\zeta^4$ and $\zeta^5$ are interchanged, the duality
of  two theories is the correspondence of $\psi$ and $-\psi$.
Moreover, the duality can also be considered as the correspondence
of $x_\mu$ and $p_\mu$ in addition to $a$ and $R$ in the complete
Yang model. Thus, this correspondence is just the duality given by
the ${\bf Z}_2:=\{e,\tau\}$ with (\ref{duality}).


By studying the motion of fictitious particle in a dimensionless
\dS$_5$-space with the radius $R/a$, we obtain a  complete Yang
model at classical level in a slightly different way under Dirac
bracket. The canonical quantization leads to the \omits{\blue
complete Yang}model at quantum level. We also re-examine why the
complete Yang model  contains both Snyder's model, the \dSR, and
their duality as a UV-IR duality\cite{YMDual}.

{\it Acknowledgements.} 
We would like to thank
  Y. Tian, K. Wu, X. N. Wu, Y. Yang and B. Zhou for valuable
  discussions.

\end{document}